\documentclass[prd,aps,preprintnumbers,showpacs,nofootinbib,superscriptaddress,notitlepage]{revtex4-1}
\pdfoutput=1

\usepackage{epsfig}
\usepackage{bm}
\usepackage{amssymb}
\usepackage{amsmath}
\usepackage{color}
\usepackage{subfigure}
\usepackage[colorlinks,
            linkcolor=blue,
            anchorcolor=black,
            citecolor=blue
            ]{hyperref}

\allowdisplaybreaks[4]

\begin{document}

\title{Probing parton saturation with forward $Z^0$-boson production at small transverse momentum in p+p and p+A collisions}

\author{Cyrille Marquet}
\affiliation{CPHT, CNRS, Ecole Polytechnique, Institut Polytechnique de Paris, Route de Saclay, 91128 Palaiseau, France}

\author{Shu-Yi Wei}
\affiliation{CPHT, CNRS, Ecole Polytechnique, Institut Polytechnique de Paris, Route de Saclay, 91128 Palaiseau, France}

\author{Bo-Wen Xiao}
\affiliation{Key Laboratory of Quark and Lepton Physics (MOE) and Institute
of Particle Physics, Central China Normal University, Wuhan 430079, China}

\begin{abstract}

We calculate and compare the differential cross sections for forward $Z^0$-boson production at small transverse momentum, in proton-proton and proton-nucleus collisions, using both the collinear and dilute-dense factorization frameworks. In both cases, we implement a Sudakov resummation of the large logarithms generated by soft-gluon emissions, which is essential in order to describe the transverse momentum distribution of forward $Z^0$ bosons measured at the Tevatron and the LHC. We further compute the nuclear modification factor in the dilute-dense framework, hoping to single out signals of saturation effects at small values of $x$. Our predictions are compared with those obtained in the collinear factorization framework, using two different nuclear parton distribution functions.

\end{abstract}
\maketitle

\section{Introduction}

The Large Hadron Collider (LHC) offers unique opportunities to study the QCD dynamics at low partonic longitudinal momentum fractions $x$. Of particular interest are observables involving particle production in the forward region, which allow to probe parton densities in one of the colliding hadrons down to $x\sim10^{-5}$. On theoretical grounds, at such low values, one expects small-$x$ effects to be relevant, in particular the high-gluon-density phenomenon known as the gluon saturation \cite{Gribov:1984tu, Mueller:1985wy, McLerran:1993ni,Gelis:2010nm}. These asymmetric processes can be described in a hybrid approach \cite{Dumitru:2005gt,Altinoluk:2011qy,Chirilli:2011km}, in which the well-understood projectile, made of a dilute parton content, is described in terms of standard collinear parton distribution functions (PDFs), while the small-$x$ dynamics in the saturated target wave function is dealt with using the Color Glass Condensate (CGC) effective theory. Then, comparing proton-proton (p+p) collisions and proton-nucleus (p+A) collisions, one can predict systematic features of non-linear QCD dynamics at high parton densities, and look for them in experiments.

As a matter of fact, one of the most convincing evidence of saturation is the suppression of the away-side peak of di-hadron azimuthal correlations in p+A versus p+p collisions \cite{Marquet:2007vb,Braidot:2010zh,Albacete:2010pg,Adare:2011sc,Stasto:2011ru,Lappi:2012nh,Stasto:2018rci,Albacete:2018ruq}, due to the bigger target saturation momentum $Q_s$ in the p+A case. Recently, those calculations have been extended to the case of forward di-jets at the LHC, using an approximation of CGC expressions suitable for high-$p_T$ particles, dubbed improved transverse momentum dependent (TMD) factorization \cite{Kotko:2015ura,vanHameren:2016ftb,vanHameren:2019ysa}. In this context, saturation effects are important when the total momentum of the di-jet system $q_T$ is of the order of $Q_s$, and the first experimental measurements released recently \cite{Aaboud:2019oop} are compatible with that narrative. 

In this kinematical regime however, the cross-section also receives important contributions from large logarithms $\ln^2\frac{p_T^2}{q_T^2}$, that could reduce the impact of saturation effects. On the theoretical side, we know how to take into account the effects of the soft-gluon emissions responsible for those large logarithms. They affect the predictive power of the fixed-order perturbative expansion, and their resummation is therefore essential. The transverse momentum resummation formalism within the collinear factorization framework (C.F.) has been established for various processes, such as semi-inclusive deep inelastic scatterings, Drell-Yan vector boson productions, and di-jet productions in hadronic collisions \cite{Collins:1981uk,Collins:1981va,Collins:1984kg,Ji:2004wu,Ji:2004xq,Sun:2014gfa,Sun:2015doa} in the past decades and has received great success in phenomenology. In particular, the transverse momentum resummation formalism gives good quantitative descriptions of transverse momentum distribution of heavy boson productions in the low-$q_T$ region at mid-rapidity \cite{Ladinsky:1993zn,Ellis:1997sc,Qiu:2000ga,Qiu:2000hf,Landry:2002ix,Bozzi:2008bb,Bozzi:2010xn,Hautmann:2012sh,Sun:2013hua,Peng:2014hta,Catani:2015vma,Bizon:2018foh,Blanco:2019qbm}. Recently, it has been proven that such Sudakov resummation can be carried out together with the small-$x$ resummation \cite{Mueller:2012uf,Mueller:2013wwa}. On the phenomenological side however, the interplay between the Sudakov and the small-$x$ logarithms has only been studied in the context of low-$p_T$ di-hadrons \cite{Stasto:2018rci}, for which non-perturbative contributions dominate and prevent robust conclusions.

Before tackling the case of forward di-jets, we consider in this letter a simpler process, namely low-$q_T$ $Z^0$-boson production, in order to study, for the first time, this interplay between Sudakov factors and non-linear small-$x$ effects, and to assess to what extend saturation effects survive the soft-gluon emissions. We find that the suppression of the low-$q_T$ $Z^0$-boson production in p+A versus p+p collisions is still present, but it gets significantly reduced. For completeness, we also compare our predictions with calculations obtained in the collinear factorization approach together with transverse momentum resummation, which can serve as a baseline without any small-$x$ saturation effects. For that process, the largely-unknown collinear nuclear gluon PDFs at small-$x$ play a key role, and as a result, the cross-section suffers from large uncertainties. Consequently, a measurement of the nuclear modification factor in agreement with our saturation prediction would also be compatible with collinear factorization with nuclear PDFs. Obviously, that conclusion could change drastically if precise small-$x$ nuclear gluon distribution were to become available.

The plan of the letter is as follows. In section II we recall the calculation of the differential cross section for $Z^0$-boson production in both the collinear factorization and CGC frameworks. In section III, we show our numerical results for the differential cross sections, compare with the available data, and give our predictions for the nuclear modification factor at forward rapidity. Section IV is devoted to conclusions.

\section{Theoretical frameworks}

In our following calculations of $Z^0$-boson production in proton-proton and proton-nucleus collisions, we adopt both the collinear and small-$x$ framework together with the Sudakov type of transverse momentum resummation, and thus we can gain some quantitative insights into the onset of small-$x$ effects through the comparison. These two frameworks are described in detail as follows. 

\subsection{Collinear factorization}

In the transverse momentum resummation (also known as the Collins-Soper-Sterman resummation) formalism \cite{Collins:1984kg}, the differential cross section of $pp\to Z^0 + X$ in the small transverse momentum region can be written as
\begin{align}
\frac{d\sigma}{2q_T dq_T}
= ~ &
\frac{4\pi^3 \alpha_{\rm em}}{3S} \int dy \int \frac{d^2b}{(2\pi)^2} e^{-i \vec q_T \cdot \vec b_\perp}
Q^2_{f \bar f} e^{-S_{\rm sud} (Q,b)} 
\nonumber \\
& 
\sum_{a,b,f} 
\int \frac{d\xi_1}{\xi_1} C_{f (\bar f),a}(x_1/\xi_1) f_{a/A}(\xi_1,\mu_b) 
\int \frac{d\xi_2}{\xi_2} C_{\bar f (f),b}(x_2/\xi_2) f_{b/B}(\xi_2,\mu_b),
\label{Eq:x-cf}
\end{align}
where the coupling is given by $Q^2_{f \bar f} = \frac{[1-4|e_f| \sin^2\theta_W]^2 + 1}{16 \sin^2 \theta_W \cos^2 \theta_W}$ with $\theta_W$ the Weinberg angle. The kinematic variables are defined as follows $Q^2 = q_T^2 + M_Z^2$, $x_1 = Q e^{y}/\sqrt{S}$ and $x_2 = Q e^{-y}/\sqrt{S}$ with $y$ the rapidity of the produced $Z^0$ boson. $f_{a/A}(\xi_1,\mu_b)$ ($f_{b/B}(\xi_2,\mu_b)$) represents the collinear parton distribution function of parton $a$ ($b$) inside the hadron $A$ ($B$). For the purpose of numerical evaluations, we choose the parametrization of PDFs $f_{a/A}$ provided by CTEQ14 \cite{Dulat:2015mca}. The Sudakov factor consists of two parts which read $S_{\rm sud} (Q,b) = S_{\rm p} (Q,b) + S_{\rm np} (Q,b)$. At the next-to-leading logarithm (NLL) accuracy, the perturbative Sudakov factor $S_{\rm p} (Q,b)$ is given by
\begin{align}
S_{\rm p} (Q,b) = \int_{\mu_b^2}^{Q^2} \frac{d\mu^2}{\mu^2} \left[ 2 \ln \frac{Q^2}{\mu^2} \left( A_1 + A_2 \right) + 2 B_1 \right],
\end{align}
where the coefficients $A_1 = C_F \frac{\alpha_s (\mu)}{2\pi}$ and $B_1 = - \frac{3}{2} C_F \frac{\alpha_s (\mu)}{2\pi}$ are derived from the one-loop correction, while $A_2 = C_F K \frac{\alpha_s^2 (\mu)}{(2\pi)^2}$, with $K= (\frac{67}{18}-\frac{\pi^2}{6}) C_A - \frac{10}{9}N_f T_R$, is computed from the two loop corrections of the double logarithmic term \cite{Kodaira:1981nh}. The QCD color factors are $C_F = 4/3$, $C_A = 3$ and $T_R = 1/2$. Here we also follow the $b_*$-prescription, which defines $b_* = b_\perp /\sqrt{1+b_\perp^2/b_{\rm max}^2}$ with $b_{\rm max} = 0.5$ GeV$^{-1}$, and $\mu_b=2e^{-\gamma_E}/b_*$. This ensures that the integration over $b_\perp$ always stays in the perturbative region for the perturbative Sudakov factor. The non-perturbative Sudakov factor can be taken from BLNY parameterization \cite{Landry:2002ix,Su:2014wpa,Prokudin:2015ysa} and it reads
\begin{align}
S_{\rm np} = g_1 b_\perp^2 + g_2 \ln\frac{Q}{Q_0} b_\perp^2 + g_3 b_\perp^2 \ln(100x_1 x_2), 
\end{align}
where the fitted parameters are found to be $g_1 = 0.21$ GeV$^2$, $g_2=0.68$ GeV$^2$, $g_3=-0.12$ GeV$^2$.

In this transverse momentum resummation formalism, $C_{i,a}(z,\mu_b) = \sum_{n=0} C_{i,a}^{(n)} (z) (\frac{\alpha_s (\mu_b)}{\pi})^{n}$, is the coefficient function to find parton $i$ inside of parton $a$ ($i,j$ represent the flavor of quark. $a,b$ could be either a quark or a gluon.), which can be calculated perturbatively \cite{Collins:1984kg, Davies:1984sp, Qiu:2000hf}.
\begin{align}
& C_{i,j}^{(0)} (z) = \delta_{ij} \delta (z-1),\quad  C_{i,g}^{(0)} (z) = 0, \\
& C_{i,j}^{(1)} (z) = \delta_{ij} \left( \frac{2}{3} (1-z) + \delta (z-1) (\frac{\pi^2}{3} - \frac{8}{3}) \right), \quad C_{i,g}^{(1)} (z) = \frac{1}{2} z (1-z). 
\end{align}
These coefficient functions take this simple form mainly because we have set the scale $\mu=\mu_b$ as in Ref. \cite{Qiu:2000hf} and those terms proportional to $\ln \mu/\mu_b$ vanish.

\subsection{Dilute-dense factorization}

\begin{figure}[h!]
\includegraphics[width=0.5\textwidth]{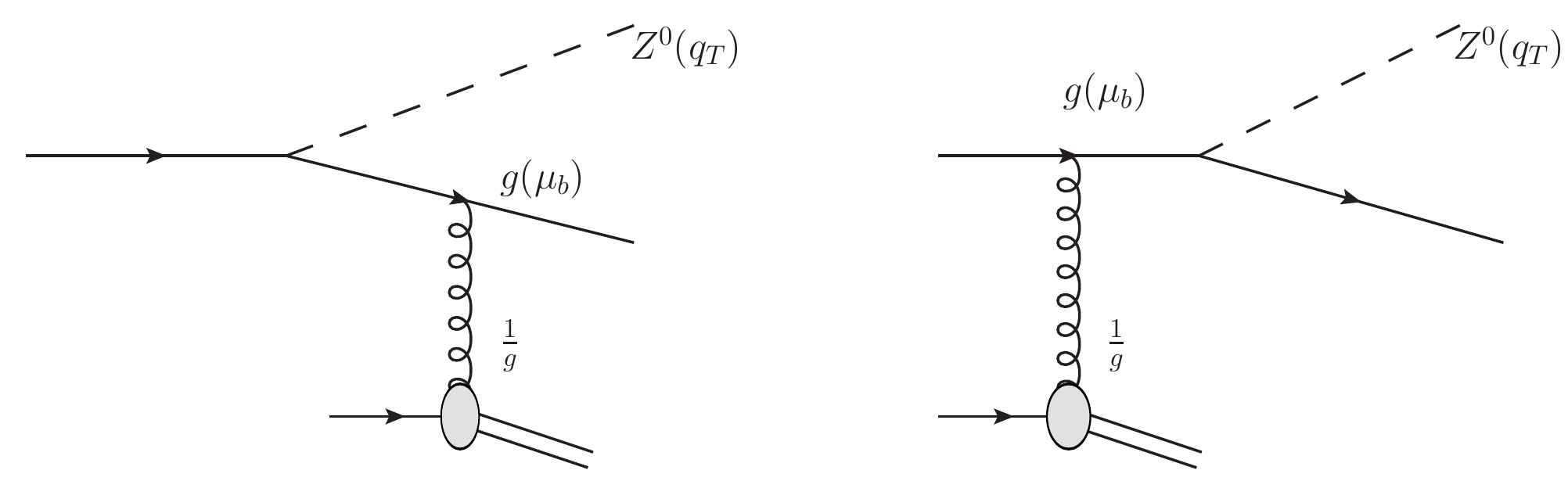}
\caption{Dominant diagrams contributing to the unintegrated quark/antiquark distribution function in the Drell-Yan process or $Z^0$ boson production in the CGC framework. These diagrams are drwan with Jaxodraw \cite{Binosi:2003yf,Binosi:2008ig}.}
\label{diagram:cgc}
\end{figure}

The dilute-dense framework uses the collinear PDF for the dilute projectile and the small-$x$ gluon distributions for the dense target. The Sudakov resummation within the dilute-dense factorization framework at one-loop level has been demonstrated in \cite{Mueller:2012uf,Mueller:2013wwa}. It is the ideal framework to compute the $Z^0$ boson production in the forward rapidity at small-$q_T$. In contrast with the di-hadron production in the forward rapidity\cite{Stasto:2018rci}, the non-perturbative effects are minimized, since fragmentation function is not involved in this process. The cross section in the dilute-dense factorization framework is given by
\begin{align}
\frac{d\sigma}{2q_T dq_T}
= 
\frac{4\pi^3 \alpha_{\rm em}}{3S} \int dy \int \frac{d^2b}{(2\pi)^2} e^{-i \vec q_T \cdot \vec b_\perp}
\sum_{f} Q^2_{f \bar f} f_{f/\bar f}(x_1,\mu_b) \mathcal{F}_{\bar f/f}(x_2,b) e^{-S'_{\rm sud} (Q,b)}.
\label{Eq:x-cgc}
\end{align}
Again the Sudakov factor $S'_{\rm sud} (Q,b)$ consists of two parts. The perturbative Sudakov factor is slightly different with that in the collinear factorization framework. The single logarithm term associate with the small-$x$ target parton does not appear in the one-loop CGC calculation. Therefore only half of the $B$-term contributes to the perturbative Sudakov factor. Due to lack of the two-loop calculation, the corresponding $A_2$-term in the dilute-dense factorization is not yet known. In this sense, our calculation in the dilute-dense factorization is only at the leading logarithm accuracy. Therefore, the perturbative Sudakov factor at one-loop level reads
\begin{align}
S'_{\rm p} (Q,b) = \int_{\mu_b^2}^{Q^2} \frac{d\mu^2}{\mu^2} \left[ 2 \ln \frac{Q^2}{\mu^2} A_1 + B_1 \right].
\end{align}
The non-perturbative Sudakov factor in the CGC framework, in principle, could be different with that in the collinear factorization framework. It can only be extracted from a global analysis to the experimental data. In this paper, we simply employ the same parameterization in the collinear factorization for simplicity.

$\mathcal{F}_{\bar f/f}(x_2,b)$, which can be computed from graphs as shown in Fig. \ref{diagram:cgc}, is the small-$x$ unintegrated quark distribution in the coordinate space, which is similar with those derived in the DIS, semi-inclusive DIS and Drell-Yan processes \cite{Mueller:1999wm,Marquet:2009ca,Xiao:2010sa,Kopeliovich:2000fb,Baier:2004tj,Gelis:2002nn,Gelis:2002fw,Gelis:2006hy,GolecBiernat:2010de,Stasto:2012ru} in the CGC framework. In coordinate space, this quark distribution reads
\begin{align}
x_2 \mathcal{F} (x_2,b) = ~& \frac{\alpha_s(\mu_b) M_Z^2 N_C}{8\pi^4 \alpha_s} \int dz d^2b_1 d^2R_\perp \frac{\vec b_1 \cdot \vec b_2}{|\vec b_1||\vec b_2|} \epsilon_f^2 K_1(\epsilon_f |\vec b_1|) K_1(\epsilon_f |\vec b_2|) \frac{1+(1-z)^2}{z} \nonumber \\
& [\mathcal{N}(x_2,z|b_1|)+\mathcal{N}(x_2,z|b_2|)-\mathcal{N}(x_2,z|b|)],
\label{eq:definition-quark-distribution}
\end{align}
where $\epsilon_f^2 = (1-z)M_Z^2$, $\vec b_2 = \vec b_1 - \vec b$ and $K_1$ is the modified Bessel function of the second kind. Phenomenologically, we use $\int d^2R_\perp =$ 15 mb for the proton target and the rcBK solution \cite{Albacete:2010sy} for the scattering amplitude $\mathcal{N}(x_2,|b|)$. There are two $\alpha_s$'s in Eq. \ref{eq:definition-quark-distribution}. The scale dependences can be different, although it is not quite obvious which scales should be used. In the phenomenological study, there are in general two prescriptions. (A) The $\alpha_s$ in the numerator comes from the $g\to q \bar q$ process and therefore its scale is chosen as $\mu_b$. However, the $\alpha_s$ in the denominator comes from the definition of unintegrated gluon distribution function in the CGC framework \cite{Dominguez:2011wm}, and its scale is related with the saturation scale. In our numerical calculation, we set the $\alpha_s$ in the denominator to be 0.3. (B) Assume the scale dependences are the same, therefore they cancel each other. We find that prescription (B) fails to describe the experimental data. It is also worth noting that more extensive discussions on this issue can be found in Refs.~\cite{Kovchegov:2007vf, Horowitz:2010yg}. We will return to this point in the next section.

\subsection{$\phi^*$ distribution}

In the experiments, the precision of the low-$q_T$ spectrum of the $Z^0$-boson is limited by the energy resolution of the detectors \cite{Banfi:2010cf,Banfi:2012du,Aad:2012wfa}. The measurement with a finer bin-size is not achievable as well for the same reason. In order to make a better constraint on the theoretical inputs with the current experimental measurements, an optimized observable, $\phi^*$, which only involves the directions of momenta, is proposed in Refs. \cite{Banfi:2010cf,Banfi:2012du}. The resolution of the $\phi^*$ measurement is supposed to be much better than the $q_T$ measurement, although the relevant physics is roughly the same.  

\begin{figure}[h!]
\includegraphics[width=0.8\textwidth]{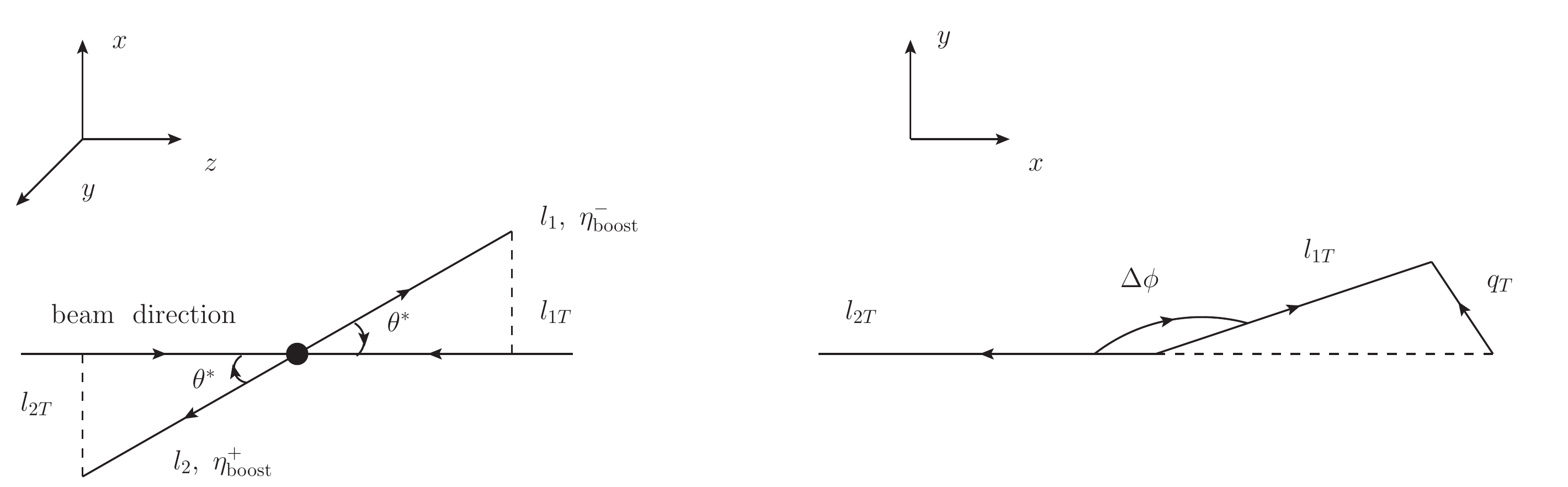}
\caption{Illustration of $\theta^*$ and $\Delta\phi$ in the definition of $\phi^*$.}
\label{fig:frame}
\end{figure}

The rapidities of the leptons are denoted by $\eta^-$ and $\eta^+$ in the lab frame. After a Lorentz boost along the beam direction, we switch to a new frame where $\eta^-_{\rm boost} = - \eta^+_{\rm boost}$. As shown in Fig. \ref{fig:frame}, $\theta^*$ is defined as the scattering angle between the lepton pair and the incoming protons in the new frame and $\Delta\phi$ is the azimuthal angle of the di-lepton pair in the transverse plane. It is straightforward to get $\cos\theta^* = \tanh\frac{\eta^--\eta^+}{2}$. $\phi^*$ is defined as,
\begin{align}
\phi^* \equiv \tan \frac{\pi-\Delta\phi}{2} \sin \theta^*.
\label{eq:definition-of-phistar}
\end{align}

In the limit $q_T \ll Q \approx M_Z$, one finds $\Delta\phi \sim \pi$ and $l_1 \approx l_2 \approx \frac{M_Z}{2}$. Furthermore, we get $\sin \theta^* \approx \frac{2l_T}{M}$ and $\phi^* = \frac{2l_T}{M_Z} \tan \frac{\pi-\Delta\phi}{2} \approx |q_T^y|/M_Z$, where $q_T^y$ is the $y$ component of $q_T$ as shown in Fig. \ref{fig:frame}. The $\phi^*$ distribution is then given by,
\begin{align}
\frac{d\sigma}{d\phi^*} 
= M_Z \frac{d\sigma}{d|q_T^y|}
= 2 M_Z \int dq_T^x \frac{d\sigma}{d^2q_T}.
\label{eq:phistar}
\end{align}

The case that $\phi^*$ is small does not in general guarantee that $q_T$ is also small, since $\phi^*$ is only determined by one component of $q_T$. However, the cross section is dominated by small-$q_T$ region. Therefore the Sudakov resummation framework is still a good approximation. From Eqs. \ref{Eq:x-cf} and \ref{Eq:x-cgc}, one can also easily derive the corresponding expressions for $\phi^*$ distribution under different frameworks by simply putting in a delta function according to the definition in Eq.~\ref{eq:definition-of-phistar}. 

At large $\phi^*$ region, $q_T$ is forced to be large. First, we would need to implement the $Y$-term \cite{Collins:1984kg,Qiu:2000hf} to give the correct calculation for $d\sigma/d^2q_T$. Second, the power correction to $\phi^* = \frac{|q_T^y|}{M} + \mathcal{O}(\frac{q_T^2}{M^2})$ becomes relevant and therefore Eq. \ref{eq:phistar} does not hold. In this paper, we only focus on the small-$q_T$ and small-$\phi^*$ region.

\section{Numerical Results}

\subsection{$Z^0$ boson production in the middle rapidity}

While the $Z^0$ boson is produced in the middle rapidity, the longitudinal momentum fractions of the incoming partons, $x_{1,2} \sim M_Z/\sqrt{S}$, are in general not small. This indicates that only the collinear factorization framework can be applied. The small-$q_T$ $Z^0$ boson cross section in the middle rapidity has been calculated numerically in several papers \cite{Ladinsky:1993zn,Ellis:1997sc,Qiu:2000ga,Qiu:2000hf,Landry:2002ix,Bozzi:2008bb,Bozzi:2010xn,Hautmann:2012sh,Sun:2013hua,Peng:2014hta,Catani:2015vma,Blanco:2019qbm} under the collinear factorization and Sudakov resummation framework. In fact, the experimental data have also been included in the global analysis to extract the parameters in the non-perturbative Sudakov factor \cite{Landry:2002ix,Su:2014wpa}. In this section, we present our numerical results in the middle rapidity and compare with the experimental data and previous studies in order to conduct a cross check.

Eq. \ref{Eq:x-cf} computes the $Z^0$ boson cross section, while, the experiments usually measure the di-lepton cross section at given rapidity window. Not all the $Z^0$ bosons produced in the middle rapidity will decay into di-lepton pairs that lie in the rapidity window covered by the detectors. In the $Z^0$ rest frame, the differential cross section of $q\bar q \to Z^0 \to l^+l^-$ is given by $d\sigma/d\cos \theta \propto A(1+\cos^2 \theta) + B \cos \theta$, with $\theta$ the angle between the outgoing lepton ($l^-$) and the incoming quark $q$. In general, both $l^-$ and $l^+$ are required to be at the same rapidity window in the experiments. The total di-lepton cross section is given by the integration of $\cos\theta$ in the full phase space, which equals the $Z^0$-boson cross section times decay fraction. The di-lepton cross section measured in the experiments is given by the $\cos\theta$-integral in the interval constrained by the rapidity window of the detectors. Therefore we get
\begin{align}
\frac{d\sigma^{\rm dilepton}}{dq_T} 
= \int_{y_{\rm min}}^{y_{\rm max}} dy_Z
\frac{d\sigma^{Z}}{dy_Z dq_T} \times 0.03366 \times \frac{3}{8} [2x + \frac{2}{3}x^3],
\label{eq:cs-dilepton}
\end{align}
where $x=\cos[2\arctan e^{-\eta}]$ with $\eta = \min [y_Z - y_{\rm min}, y_{\rm max} - y_Z]$. $y_{\rm min}$ and $y_{\rm max}$ define the rapidity window of the di-lepton pair measured in experiments. $0.03366$ is the fraction of the $Z^0$ boson decays to a certain flavor of di-lepton pair. When the rapidity window is wide, $\eta$ is large and $x \sim 1$. Most of the di-lepton pairs can be measured by the detectors. The cross section of di-lepton pair is simply that of the $Z^0$-boson multiplies the decay fraction. However, in case that the rapidity window is very narrow, this factor could give a sizable suppression.

\begin{figure}[h!]
\includegraphics[width=0.3\textwidth]{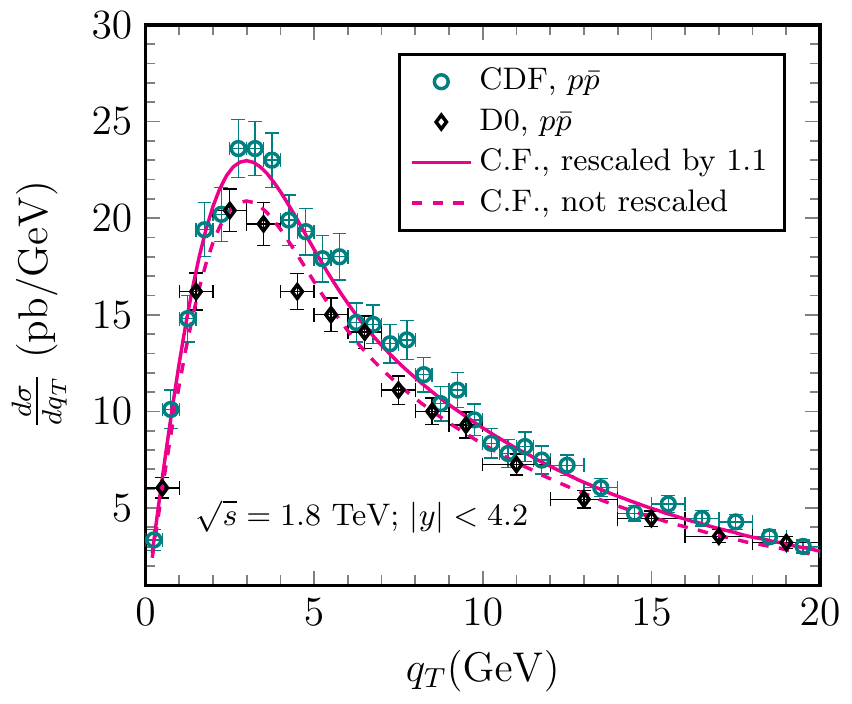}
\includegraphics[width=0.3\textwidth]{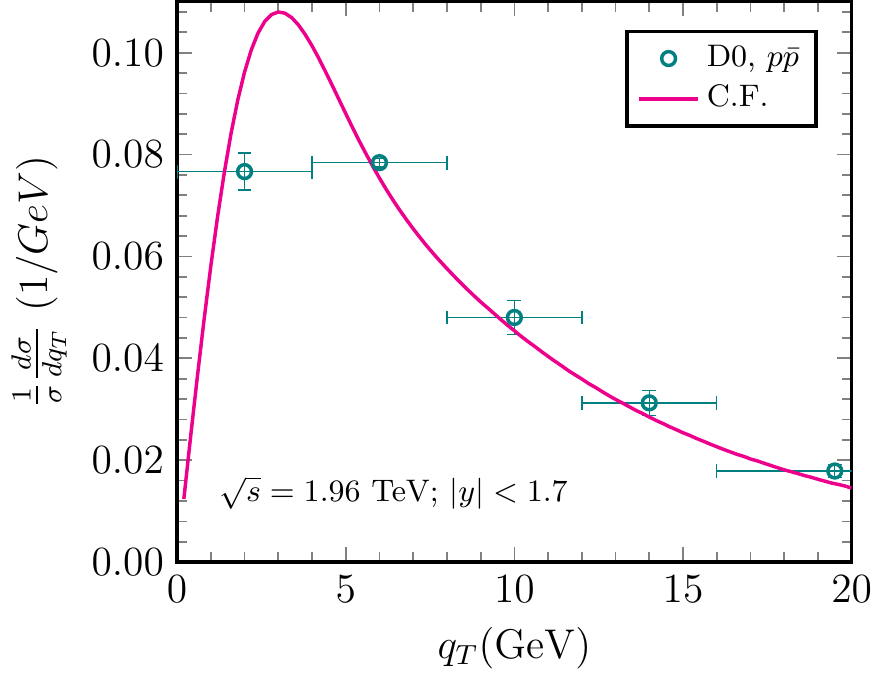}
\includegraphics[width=0.3\textwidth]{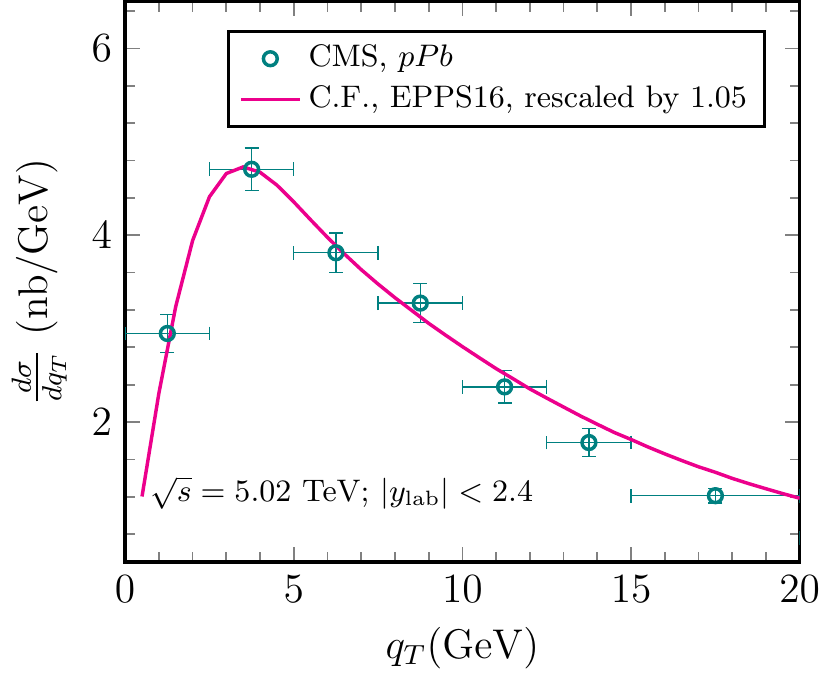}
\caption{
$q_T$ distribution of di-lepton pair (through $Z^0$ boson) compared with data from CDF collaboration \cite{Affolder:1999jh}, D0 collaboration \cite{Abbott:1999wk,Abazov:2010kn} and CMS collaboration \cite{Khachatryan:2015pzs}. 
\label{fig:qT-middle}
}
\end{figure}

We compute the differential cross section and compare with the experimental data measured by CDF \cite{Affolder:1999jh}, D0 \cite{Abbott:1999wk,Abazov:2010kn} and CMS \cite{Khachatryan:2015pzs} collaborations in Fig. \ref{fig:qT-middle}. It has already been shown in \cite{Landry:2002ix} the CDF data \cite{Affolder:1999jh} is systematically $10\%$ larger than the D0 data \cite{Abbott:1999wk}. The parameterization for the non-perturbative Sudakov factor \cite{Landry:2002ix} used by us in the numerical evaluation is determined by fitting with the D0 data. Therefore, our results is about $10\%$ smaller than the CDF data. Despite that, the theoretical results show good agreement with the experimental data. The shape of $q_T$ distribution is similar with that in Ref. \cite{Kang:2012am}, although different methods have been employed to deal with the non-perturbative effect. This confirms that the non-perturbative Sudakov factor has little impact on the shape of the $q_T$ distribution. But it may have some effects on the overall normalization.

\begin{figure}[h!]
\includegraphics[width=0.3\textwidth]{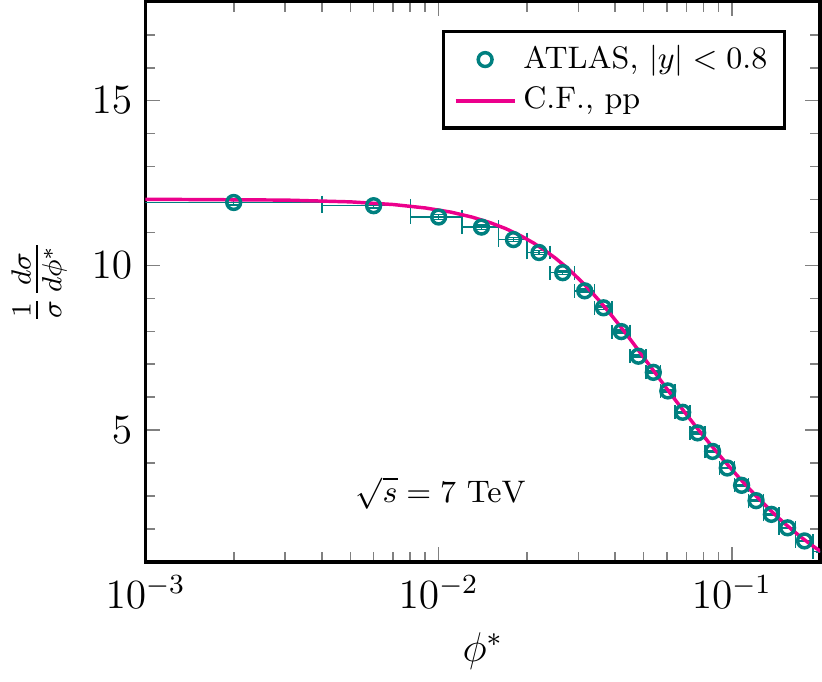}
\caption{
Self-normalized $\phi^*$ distribution of di-lepton pair (through $Z^0$ boson) compared with the ATLAS \cite{Aad:2012wfa} data.
\label{fig:phistar-atlas}
}
\end{figure}

We also calculate the self-normalized $\phi^*$ distribution in the collinear factorization framework using Eq. \ref{eq:phistar} and compare with the ATLAS data \cite{Aad:2012wfa} in Fig. \ref{fig:phistar-atlas}. As discussed in the previous section, the error band of the experimental data has been significantly reduced as compared to the $q_T$ measurement even with finer bin-sizes. The collinear framework calculation provides a good baseline description of the experimental data at small $\phi^*$.

\subsection{$Z^0$ boson production in the forward rapidity}

When the $Z^0$ boson is produced in the forward rapidity, the parton momentum fraction of the projectile side becomes large while that of the target side is small. At the LHC, $\sqrt{S} \approx 5$ TeV, $x_2$ is smaller than $10^{-2}$ when $y>1$. At Tevatron, $\sqrt{S} \approx 2$ TeV, $x_2<10^{-2}$ when $y>2$. In these cases, the projectile proton is regarded as a dilute system while the target proton or nucleus can be viewed as a dense system. The so-called dilute-dense factorization becomes the ideal framework to study the $Z^0$ production process in the forward rapidity. On the other hand, the collinear factorization, which works perfectly in the middle rapidity, can still apply, since the momentum fraction of the target parton is not yet extremely small.

\begin{figure}[htb]
\includegraphics[width=0.3\textwidth]{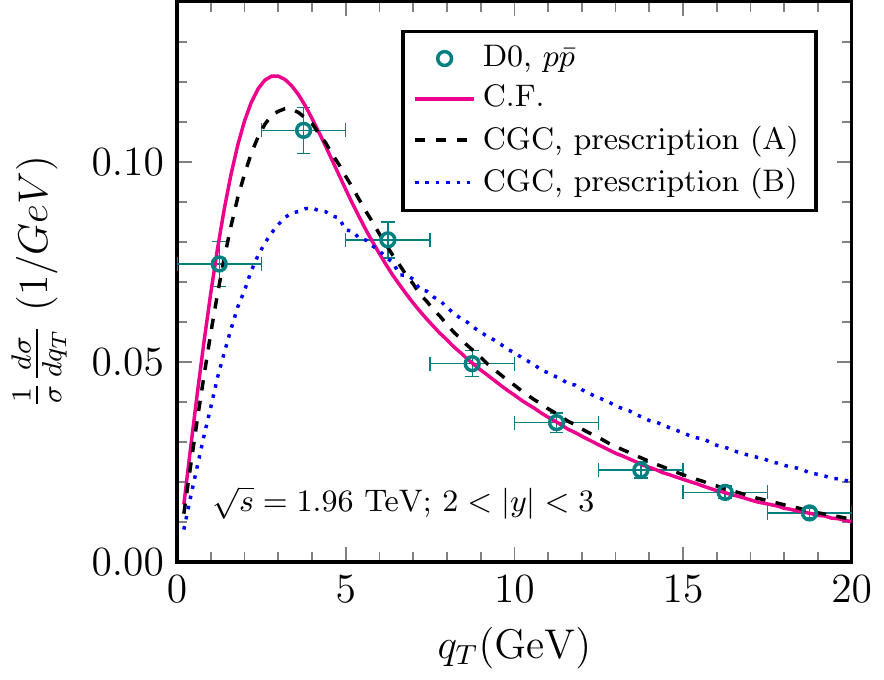}
\includegraphics[width=0.3\textwidth]{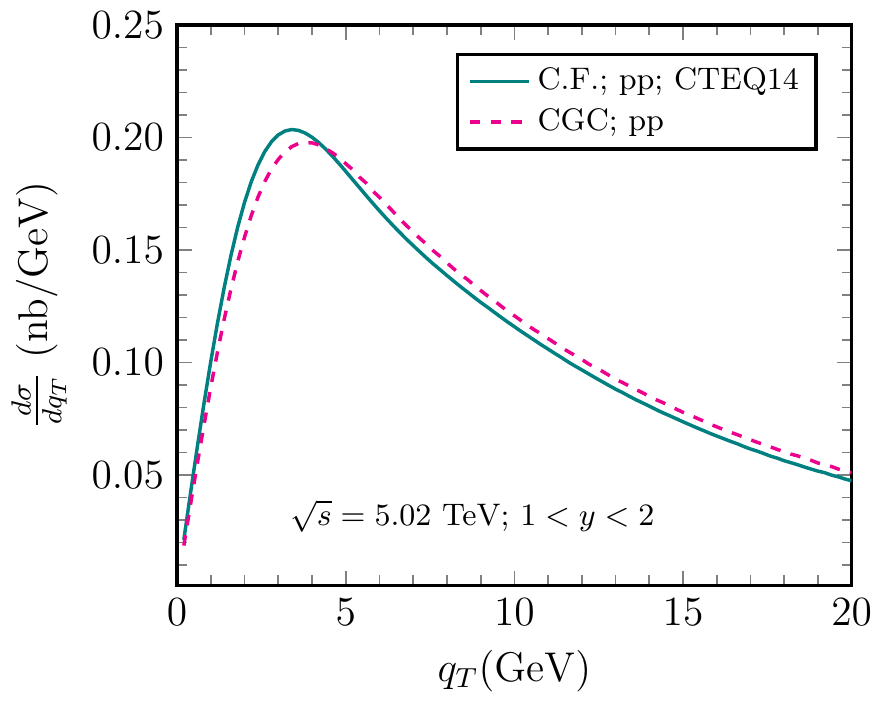}
\includegraphics[width=0.3\textwidth]{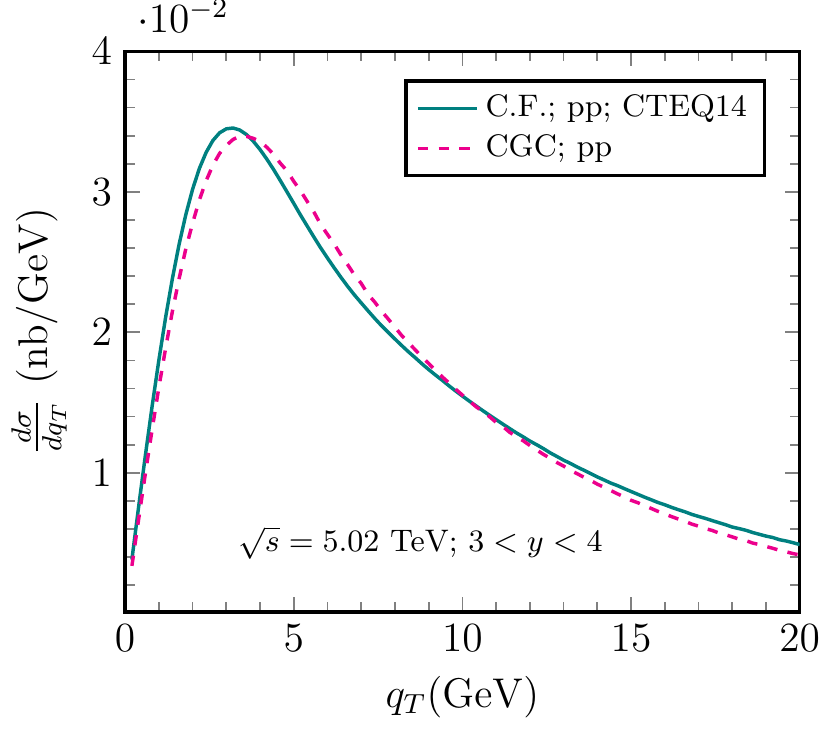}
\caption{
Left: Self-normalized $q_T$ distributions of di-lepton pair (through $Z^0$ boson) in forward $p+\bar p$ collisions at $\sqrt{S}=1.96$ TeV compared with D0 \cite{Abazov:2007ac} data. Middle and right: Predictions for $q_T$ distribution of $Z^0$ boson in forward p+p collisions at $\sqrt{S}=5.02$ TeV.
\label{fig:D0}
}
\end{figure}

In Fig. \ref{fig:D0}, we calculate the self-normalized $Z^0$ boson cross section in the forward $p\bar p$ collisions and compare with the D0 data \cite{Abazov:2007ac}. In the CGC framework, there is no difference between proton and antiproton, since all the quarks and antiquarks come from the gluon splittings (i.e., they are sea partons.). The contribution from the valence anti-quark is missing. Therefore, the cross section calculated with the CGC framework is about $20\%$ smaller than that given by the collinear factorization framework. However, this difference disappears in p+p collisions, as shown in the right two figures in Fig. \ref{fig:D0}. 

As discussed in the previous section, the leftmost figure in Fig. \ref{fig:D0} obviously shows that prescription (B) (the blue dotted curve) fails to yield the correct $q_T$ distribution, while prescription (A) along with the collinear factorization can. Therefore, it is essential to keep the different scale dependences in the couplings. In the paper, we will always adopt the prescription (A) in the numerical calculations.

\begin{figure}[h!]
\includegraphics[width=0.3\textwidth]{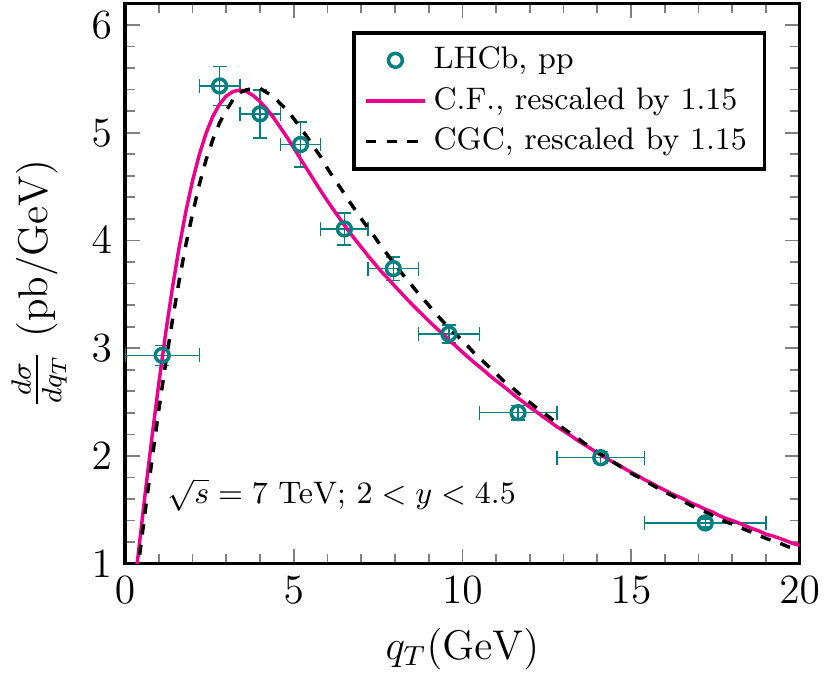}
\includegraphics[width=0.3\textwidth]{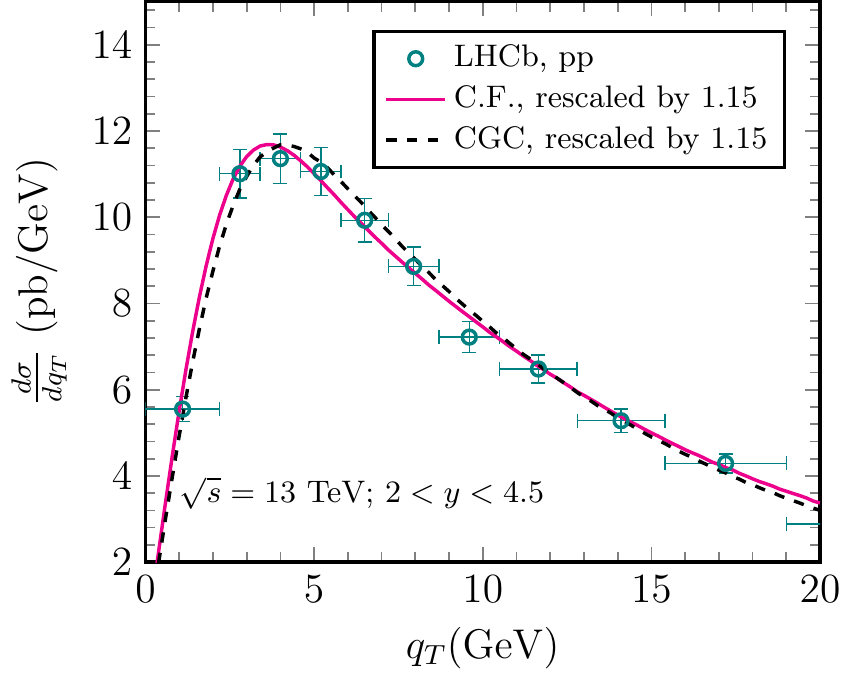}\\
\includegraphics[width=0.3\textwidth]{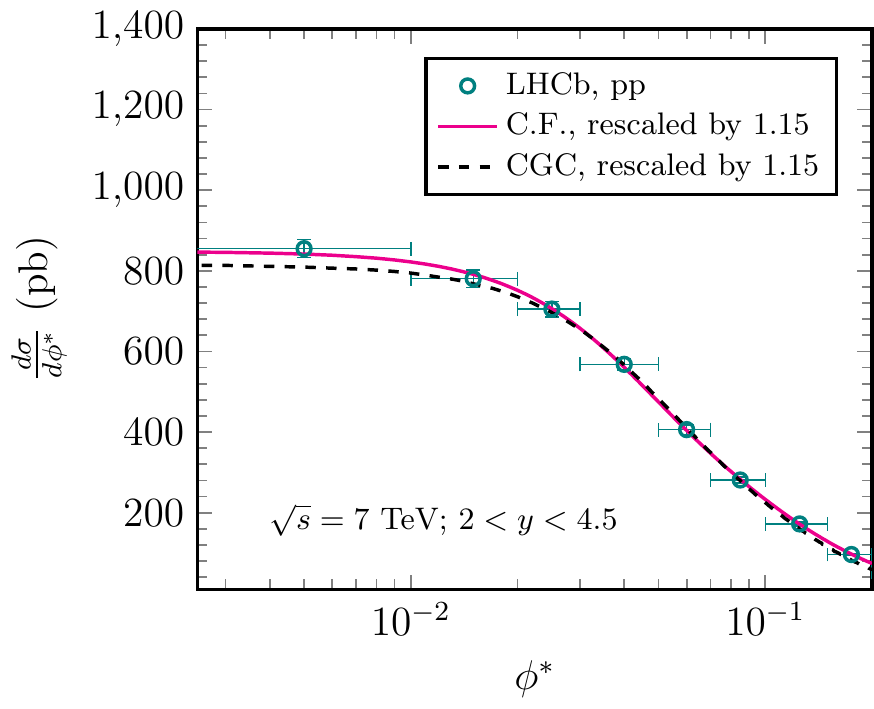}
\includegraphics[width=0.3\textwidth]{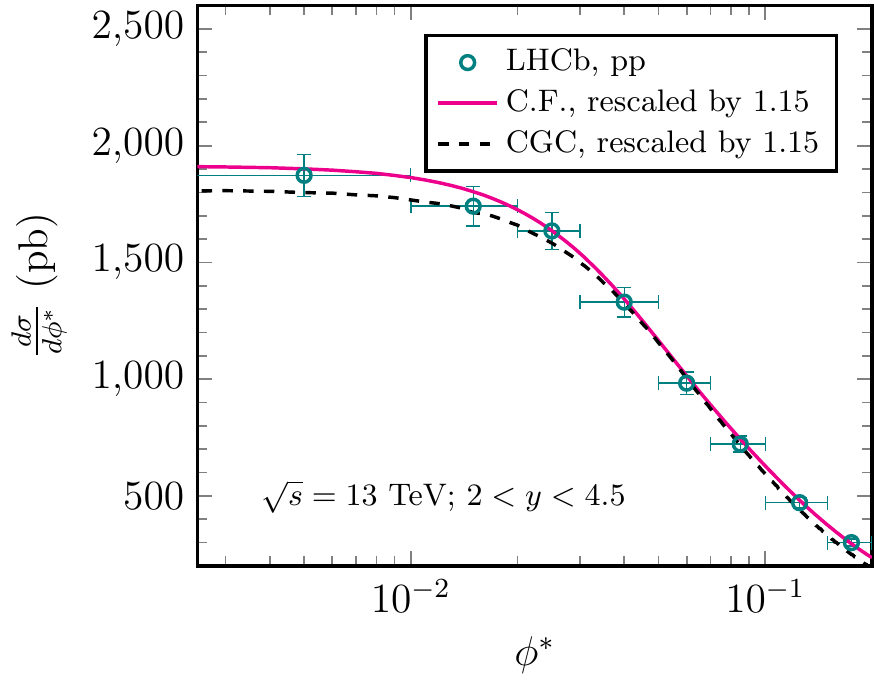}
\caption{
$q_T$ distribution of di-lepton pair (through $Z^0$ boson) in the forward p+p collisions at $\sqrt{S} =$ 7 TeV \cite{Aaij:2015gna} and 13 TeV \cite{Aaij:2016mgv} measured by the LHCb collaboration.
}
\label{fig:lhcb}
\end{figure}

The LHCb collaboration has measured the absolute differential cross section \cite{Aaij:2015gna, Aaij:2016mgv} in the forward p+p collisions as well. We compare our results with the experimental data in Fig. \ref{fig:lhcb}. Our theoretical calculation can describe the shapes of the distributions quite well, thanks to the success of the Sudakov resummation. However, we need to rescale our cross section by 1.15, which is larger than the one used in the middle rapidity, in order to match the absolute cross section. So far, we cannot understand well why a larger rescaling factor is required. However, according to Ref. \cite{Landry:2002ix}, the overall normalization factor for different experiments in the BLNY parameterization could vary from $0.86$ to $1.19$. We assume this normalization factor remains the same in p+p and p+A collisions and therefore it is canceled in the nuclear modification factor.

\begin{figure}[h!]
\includegraphics[width=0.3\textwidth]{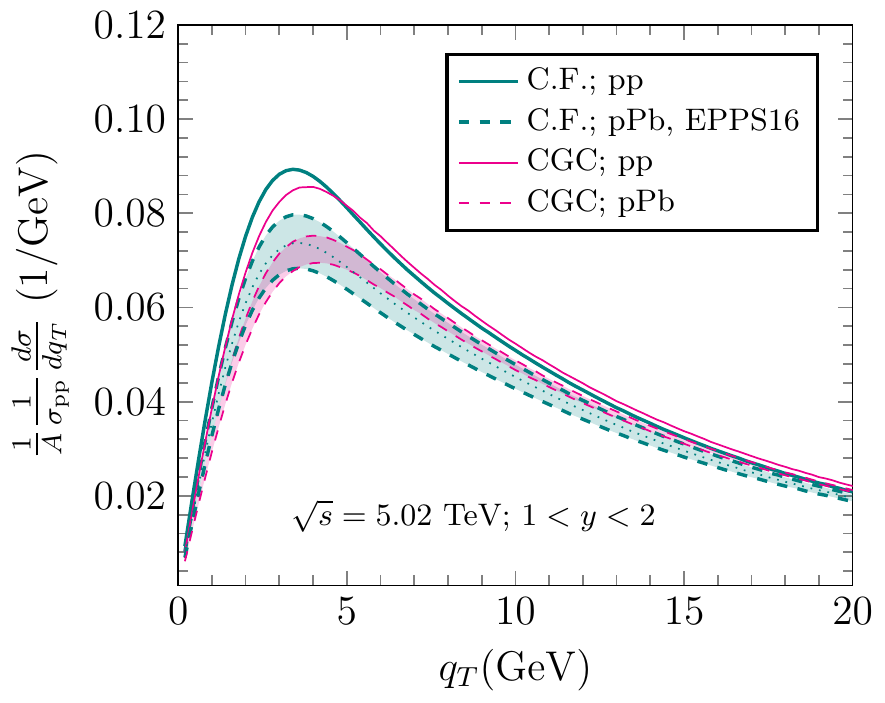}
\includegraphics[width=0.3\textwidth]{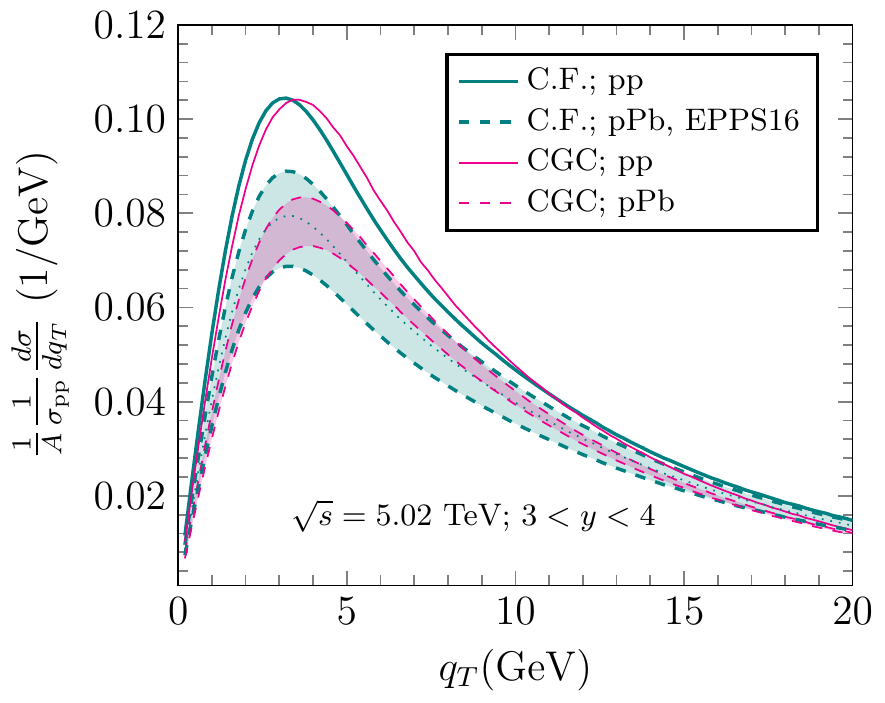} \\
\includegraphics[width=0.3\textwidth]{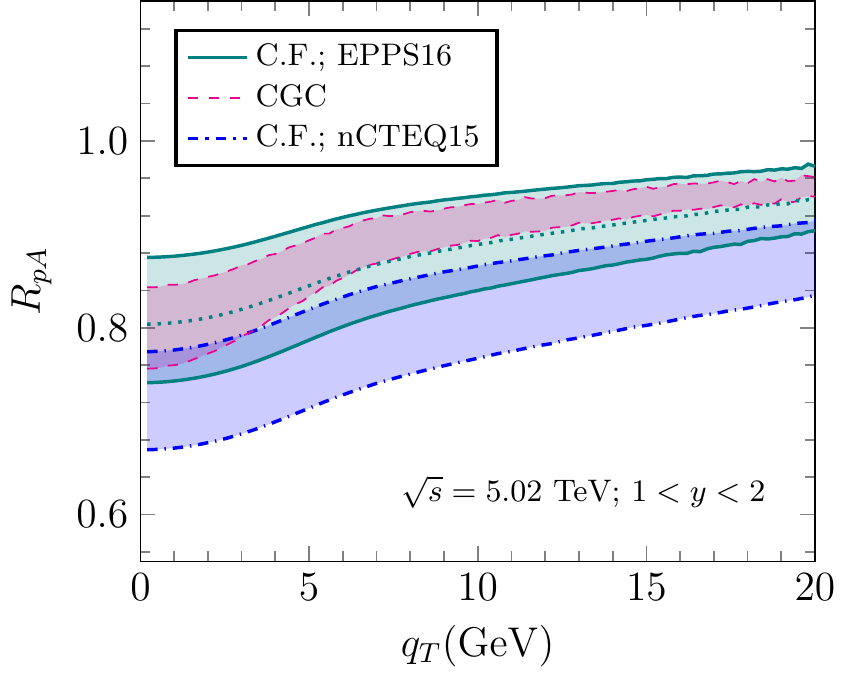}
\includegraphics[width=0.3\textwidth]{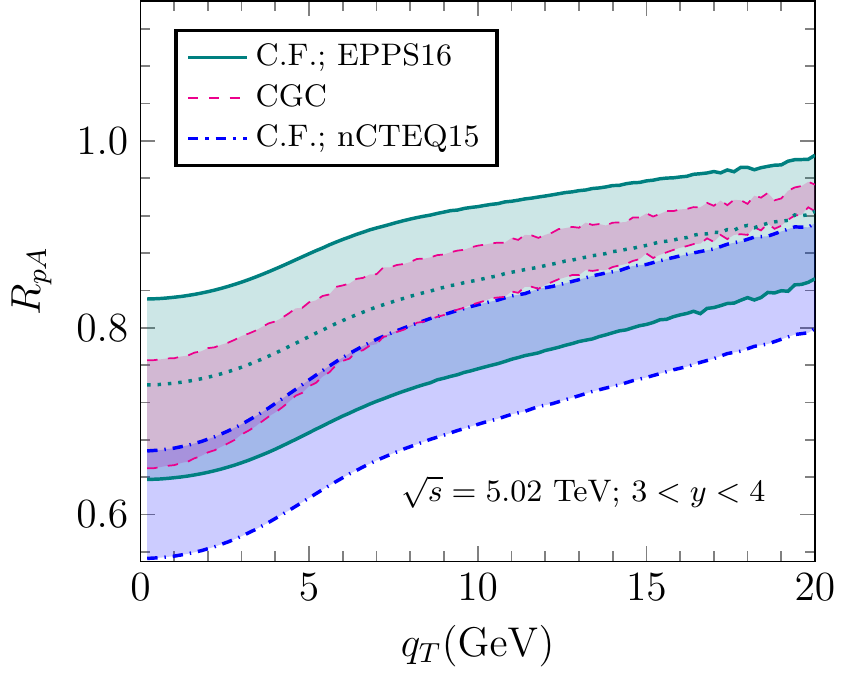}
\caption{
Normalized $q_T$ distribution of $Z^0$ boson in forward p+p and p+Pb collisions at $\sqrt{S_{NN}} =$ 5.02TeV and the corresponding nuclear suppression factor.
\label{fig:qT-pA}
}
\end{figure}

The modification of cross sections in forward p+p and p+A collisions is described by different languages in different frameworks. In the CGC framework, the gluon density is much larger in the nucleus than that in the proton and therefore the saturation effect is much stronger, thus $R_{pA}$ is smaller than one. In the collinear factorization approach, the cold nuclear effects are factorized out from the hard interaction and absorbed into the nuclear modified parton distribution functions (nPDFs). There are several parameterizations of nPDFs in the market \cite{Hirai:2007sx,deFlorian:2011fp,Kulagin:2014vsa,Kovarik:2015cma,Khanpour:2016pph,Eskola:2016oht}. In this paper, we employ the nPDFs provided by nCTEQ \cite{Kovarik:2015cma} and EPPS16 \cite{Eskola:2016oht} groups to evaluate the nuclear modification factor, which is shown in Fig. \ref{fig:qT-pA}, along with the prediction given by the CGC framework. In p+p collisions, the cross section is self normalized in the range $0<q_T<20$ GeV, while in p+A collisions, it is normalized by $\frac{1}{A} \frac{1}{\sigma_{\rm pp}} \frac{d\sigma_{\rm pA}}{dq_T}$. By doing so, we can demonstrate not only the modification of the shapes, but also the suppression of the cross sections. The ratio of the ``normalized'' differential cross sections gives the nuclear modification factor, $R_{pA} = \frac{1}{A} \frac{d\sigma_{pA}/dq_T}{d\sigma_{pp}/dq_T}$.

In collinear factorization, the error bands are given by uncertainties of nPDFs, while in the CGC framework, the error band is given by varying the saturation scale of the nucleus from $2Q_{sp}^2$ to $3Q_{sp}^2$ as the initial condition to solve the rcBK evolution equation. Fig. \ref{fig:qT-pA} shows that the CGC calculation agrees with the result given by the collinear factorization framework with EPPS16 nPDFs, albeit with a smaller error band. However, there is a clear difference between the CGC calculation and the nCTEQ15 nPDFs results. This is mainly because that the nPDFs at small-$x$ are in general less constrained due to the lack of experimental data. The future measurements of the nuclear modification factor to the $Z^0$ boson production in the forward rapidity at the LHC should be included in the global analysis and would play an important role to reduce the uncertainty of the nPDFs at small-$x$.

\begin{figure}[h!]
\includegraphics[width=0.3\textwidth]{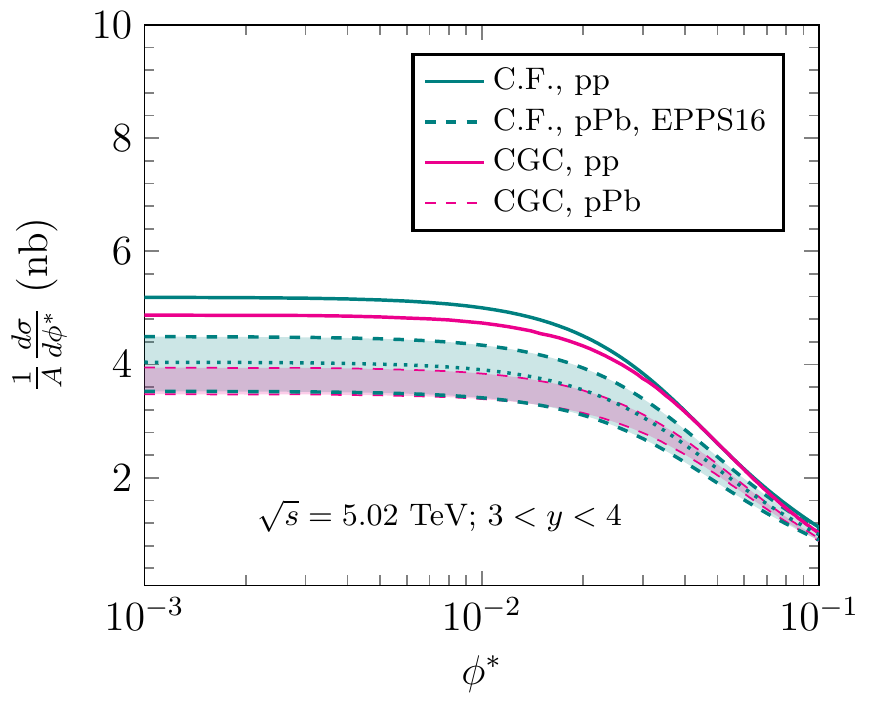}
\includegraphics[width=0.3\textwidth]{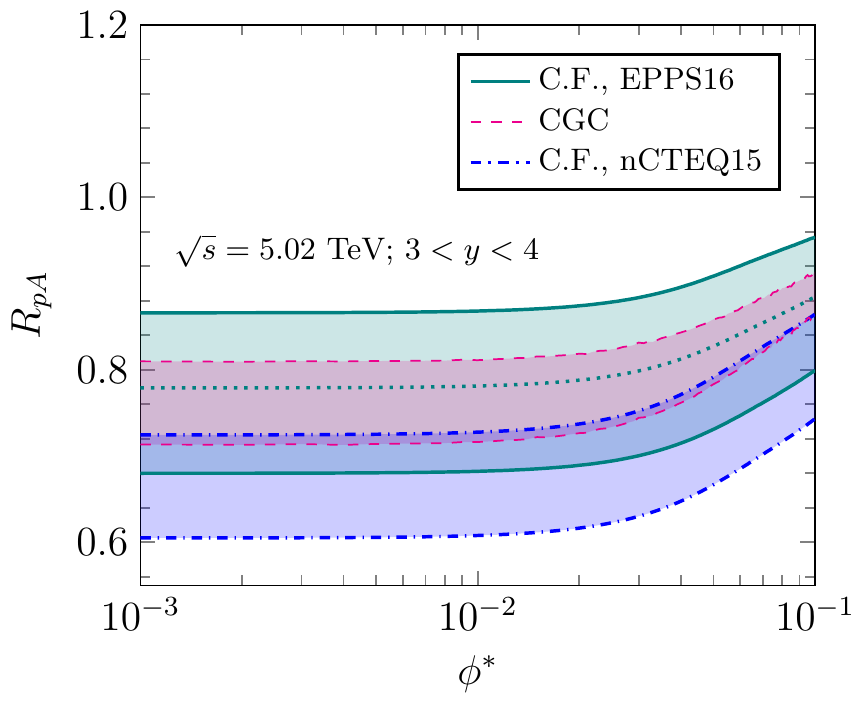}
\caption{
Nuclear number normalized $\phi^*$ distribution of $Z^0$ boson in forward p+p and p+Pb collisions at $\sqrt{S_{NN}} =$ 5.02TeV and the corresponding nuclear suppression factor.
\label{fig:phistar-pA}
}
\end{figure}

Furthermore, using Eq. \ref{eq:phistar}, we can calculate the $\phi^*$ distributions in forward p+p and p+A collisions which are shown in Fig. \ref{fig:phistar-pA}. In spite of a small difference between two frameworks in the calculation of $d\sigma/d\phi^*$, both CGC framework and collinear factorization framework with the EPPS16 nPDF predict the similar nuclear suppression factor, $R_{pA}$. In addition, we also find that CTEQ15 nPDF predicts a stronger suppression at small-$x$. 

\section{Summary}

To summarize, we calculated the differential $Z^0$ boson cross section in forward p+p and p+A collisions at small transverse momentum by applying Sudakov resummation in both collinear factorization framework and CGC framework. We find different nPDFs groups predict different nuclear modification effects and the uncertainties are very large at small-$x$. The CGC calculation coincides with the expectations given by EPPS16 nPDFs, with a relatively smaller error band. The upcoming LHC data can help to constrain the parameters in the nPDFs and improve our understanding about the dynamics of parton saturation, in particular the nonlinear evolution at small-$x$.

\begin{acknowledgments}
CM and SYW are supported by the Agence Nationale de la Recherche under the project ANR-16-CE31-0019-02, and BX is supported by the Natural Science Foundation of China (NSFC) under Grant Nos.~11575070.
\end{acknowledgments}


\begin{thebibliography}{99}


\bibitem{Gribov:1984tu} L.~V.~Gribov, E.~M.~Levin and M.~G.~Ryskin,
Phys.\ Rept.\ \textbf{100}, 1 (1983). 


\bibitem{Mueller:1985wy} A.~H.~Mueller and J.~W.~Qiu,
Nucl.\ Phys.\ B \textbf{268}, 427 (1986). 

\bibitem{McLerran:1993ni} L.~D.~McLerran and R.~Venugopalan,
Phys.\ Rev.\ D \textbf{49}, 2233 (1994); 
Phys.\ Rev.\ D \textbf{49}, 3352 (1994). 


\bibitem{Gelis:2010nm} 
  F.~Gelis, E.~Iancu, J.~Jalilian-Marian and R.~Venugopalan,
  Ann.\ Rev.\ Nucl.\ Part.\ Sci.\  {\bf 60}, 463 (2010).



\bibitem{Dumitru:2005gt} 
  A.~Dumitru, A.~Hayashigaki and J.~Jalilian-Marian,
  Nucl.\ Phys.\ A {\bf 765}, 464 (2006).



\bibitem{Altinoluk:2011qy} 
  T.~Altinoluk and A.~Kovner,
  Phys.\ Rev.\ D {\bf 83}, 105004 (2011).



\bibitem{Chirilli:2011km} 
  G.~A.~Chirilli, B.~W.~Xiao and F.~Yuan,
  Phys.\ Rev.\ Lett.\  {\bf 108}, 122301 (2012).



\bibitem{Marquet:2007vb} 
  C.~Marquet,
  Nucl.\ Phys.\ A {\bf 796}, 41 (2007).



\bibitem{Braidot:2010zh} 
  E.~Braidot [STAR Collaboration],
  arXiv:1005.2378 [hep-ph].



\bibitem{Albacete:2010pg} 
  J.~L.~Albacete and C.~Marquet,
  Phys.\ Rev.\ Lett.\  {\bf 105}, 162301 (2010).



\bibitem{Adare:2011sc} 
  A.~Adare {\it et al.} [PHENIX Collaboration],
  Phys.\ Rev.\ Lett.\  {\bf 107}, 172301 (2011).



\bibitem{Stasto:2011ru} 
  A.~Stasto, B.~W.~Xiao and F.~Yuan,
  Phys.\ Lett.\ B {\bf 716}, 430 (2012).



\bibitem{Lappi:2012nh} 
  T.~Lappi and H.~Mantysaari,
  Nucl.\ Phys.\ A {\bf 908}, 51 (2013).


\bibitem{Stasto:2018rci} 
  A.~Stasto, S.~Y.~Wei, B.~W.~Xiao and F.~Yuan,
  Phys.\ Lett.\ B {\bf 784}, 301 (2018).


\bibitem{Albacete:2018ruq}
  J.~L.~Albacete, G.~Giacalone, C.~Marquet and M.~Matas,
  Phys.\ Rev.\ D {\bf 99} (2019) no.1,  014002.
  
  

\bibitem{Kotko:2015ura} 
  P.~Kotko, K.~Kutak, C.~Marquet, E.~Petreska, S.~Sapeta and A.~van Hameren,
  JHEP {\bf 1509}, 106 (2015).



\bibitem{vanHameren:2016ftb} 
  A.~van Hameren, P.~Kotko, K.~Kutak, C.~Marquet, E.~Petreska and S.~Sapeta,
  JHEP {\bf 1612}, 034 (2016);
  Erratum: [JHEP {\bf 1902}, 158 (2019)].


\bibitem{vanHameren:2019ysa}
  A.~van Hameren, P.~Kotko, K.~Kutak and S.~Sapeta,
  Phys.\ Lett.\ B {\bf 795} 511 (2019).


\bibitem{Aaboud:2019oop} 
  M.~Aaboud {\it et al.} [ATLAS Collaboration],
  arXiv:1901.10440 [nucl-ex].



\bibitem{Collins:1981uk} 
  J.~C.~Collins and D.~E.~Soper,
  Nucl.\ Phys.\ B {\bf 193}, 381 (1981);
  Erratum: [Nucl.\ Phys.\ B {\bf 213}, 545 (1983)].



\bibitem{Collins:1981va} 
  J.~C.~Collins and D.~E.~Soper,
  Nucl.\ Phys.\ B {\bf 197}, 446 (1982).



\bibitem{Collins:1984kg} 
  J.~C.~Collins, D.~E.~Soper and G.~F.~Sterman,
  Nucl.\ Phys.\ B {\bf 250}, 199 (1985).



\bibitem{Ji:2004wu} 
  X.~d.~Ji, J.~p.~Ma and F.~Yuan,
  Phys.\ Rev.\ D {\bf 71}, 034005 (2005).



\bibitem{Ji:2004xq} 
  X.~d.~Ji, J.~P.~Ma and F.~Yuan,
  Phys.\ Lett.\ B {\bf 597}, 299 (2004).



\bibitem{Sun:2014gfa} 
  P.~Sun, C.-P.~Yuan and F.~Yuan,
  Phys.\ Rev.\ Lett.\  {\bf 113}, no. 23, 232001 (2014).



\bibitem{Sun:2015doa} 
  P.~Sun, C.-P.~Yuan and F.~Yuan,
  Phys.\ Rev.\ D {\bf 92}, no. 9, 094007 (2015).



\bibitem{Ladinsky:1993zn} 
  G.~A.~Ladinsky and C.~P.~Yuan,
  Phys.\ Rev.\ D {\bf 50}, R4239 (1994).



\bibitem{Ellis:1997sc} 
  R.~K.~Ellis, D.~A.~Ross and S.~Veseli,
  Nucl.\ Phys.\ B {\bf 503}, 309 (1997).



\bibitem{Qiu:2000ga} 
  J.~w.~Qiu and X.~f.~Zhang,
  Phys.\ Rev.\ Lett.\  {\bf 86}, 2724 (2001).



\bibitem{Qiu:2000hf} 
  J.~w.~Qiu and X.~f.~Zhang,
  Phys.\ Rev.\ D {\bf 63}, 114011 (2001).



\bibitem{Landry:2002ix} 
  F.~Landry, R.~Brock, P.~M.~Nadolsky and C.~P.~Yuan,
  Phys.\ Rev.\ D {\bf 67}, 073016 (2003).



\bibitem{Bozzi:2008bb} 
  G.~Bozzi, S.~Catani, G.~Ferrera, D.~de Florian and M.~Grazzini,
  Nucl.\ Phys.\ B {\bf 815}, 174 (2009).



\bibitem{Bozzi:2010xn} 
  G.~Bozzi, S.~Catani, G.~Ferrera, D.~de Florian and M.~Grazzini,
  Phys.\ Lett.\ B {\bf 696}, 207 (2011).



\bibitem{Hautmann:2012sh} 
  F.~Hautmann, M.~Hentschinski and H.~Jung,
  Nucl.\ Phys.\ B {\bf 865}, 54 (2012).



\bibitem{Sun:2013hua} 
  P.~Sun and F.~Yuan,
  Phys.\ Rev.\ D {\bf 88}, no. 11, 114012 (2013).



\bibitem{Peng:2014hta} 
  J.~C.~Peng and J.~W.~Qiu,
  Prog.\ Part.\ Nucl.\ Phys.\  {\bf 76}, 43 (2014).



\bibitem{Catani:2015vma} 
  S.~Catani, D.~de Florian, G.~Ferrera and M.~Grazzini,
  JHEP {\bf 1512}, 047 (2015).

\bibitem{Bizon:2018foh}
  W.~Bizon {\it et al.},
  JHEP {\bf 1812}, 132 (2018).



\bibitem{Blanco:2019qbm} 
  E.~Blanco, A.~van Hameren, H.~Jung, A.~Kusina and K.~Kutak,
  arXiv:1905.07331 [hep-ph].



\bibitem{Mueller:2012uf} 
  A.~H.~Mueller, B.~W.~Xiao and F.~Yuan,
  Phys.\ Rev.\ Lett.\  {\bf 110}, no. 8, 082301 (2013).



\bibitem{Mueller:2013wwa} 
  A.~H.~Mueller, B.~W.~Xiao and F.~Yuan,
  Phys.\ Rev.\ D {\bf 88}, no. 11, 114010 (2013).




\bibitem{Dulat:2015mca} 
  S.~Dulat {\it et al.},
  Phys.\ Rev.\ D {\bf 93}, no. 3, 033006 (2016).



\bibitem{Kodaira:1981nh} 
  J.~Kodaira and L.~Trentadue,
  Phys.\ Lett.\  {\bf 112B}, 66 (1982).



\bibitem{Su:2014wpa} 
  P.~Sun, J.~Isaacson, C.-P.~Yuan and F.~Yuan,
  Int.\ J.\ Mod.\ Phys.\ A {\bf 33}, no. 11, 1841006 (2018).



\bibitem{Prokudin:2015ysa} 
  A.~Prokudin, P.~Sun and F.~Yuan,
  Phys.\ Lett.\ B {\bf 750}, 533 (2015).



\bibitem{Davies:1984sp} 
  C.~T.~H.~Davies, B.~R.~Webber and W.~J.~Stirling,
  Nucl.\ Phys.\ B {\bf 256}, 413 (1985).

\bibitem{Binosi:2003yf} 
  D.~Binosi and L.~Theussl,
  Comput.\ Phys.\ Commun.\  {\bf 161}, 76 (2004).

\bibitem{Binosi:2008ig} 
  D.~Binosi, J.~Collins, C.~Kaufhold and L.~Theussl,
  Comput.\ Phys.\ Commun.\  {\bf 180}, 1709 (2009).

\bibitem{Mueller:1999wm} 
  A.~H.~Mueller,
  Nucl.\ Phys.\ B {\bf 558}, 285 (1999).



\bibitem{Marquet:2009ca} 
  C.~Marquet, B.~W.~Xiao and F.~Yuan,
  Phys.\ Lett.\ B {\bf 682}, 207 (2009).



\bibitem{Xiao:2010sa} 
  B.~W.~Xiao and F.~Yuan,
  Phys.\ Rev.\ D {\bf 82}, 114009 (2010).



\bibitem{Kopeliovich:2000fb} 
  B.~Z.~Kopeliovich, J.~Raufeisen and A.~V.~Tarasov,
  Phys.\ Lett.\ B {\bf 503}, 91 (2001).



\bibitem{Baier:2004tj} 
  R.~Baier, A.~H.~Mueller and D.~Schiff,
  Nucl.\ Phys.\ A {\bf 741}, 358 (2004).



\bibitem{Gelis:2002nn} 
  F.~Gelis and J.~Jalilian-Marian,
  Phys.\ Rev.\ D {\bf 67}, 074019 (2003).



\bibitem{Gelis:2002fw} 
  F.~Gelis and J.~Jalilian-Marian,
  Phys.\ Rev.\ D {\bf 66}, 094014 (2002).



\bibitem{Gelis:2006hy} 
  F.~Gelis and J.~Jalilian-Marian,
  Phys.\ Rev.\ D {\bf 76}, 074015 (2007).



\bibitem{GolecBiernat:2010de} 
  K.~Golec-Biernat, E.~Lewandowska and A.~M.~Stasto,
  Phys.\ Rev.\ D {\bf 82}, 094010 (2010).



\bibitem{Stasto:2012ru} 
  A.~Stasto, B.~W.~Xiao and D.~Zaslavsky,
  Phys.\ Rev.\ D {\bf 86}, 014009 (2012).



\bibitem{Albacete:2010sy} 
  J.~L.~Albacete, N.~Armesto, J.~G.~Milhano, P.~Quiroga-Arias and C.~A.~Salgado,
  Eur.\ Phys.\ J.\ C {\bf 71}, 1705 (2011).



\bibitem{Dominguez:2011wm} 
  F.~Dominguez, C.~Marquet, B.~W.~Xiao and F.~Yuan,
  Phys.\ Rev.\ D {\bf 83}, 105005 (2011).


\bibitem{Kovchegov:2007vf}
  Y.~V.~Kovchegov and H.~Weigert,
  Nucl.\ Phys.\ A {\bf 807} (2008) 158.
  
%
\bibitem{Horowitz:2010yg} 
  W.~A.~Horowitz and Y.~V.~Kovchegov,
  Nucl.\ Phys.\ A {\bf 849}, 72 (2011).
  

\bibitem{Banfi:2010cf} 
  A.~Banfi, S.~Redford, M.~Vesterinen, P.~Waller and T.~R.~Wyatt,
  Eur.\ Phys.\ J.\ C {\bf 71}, 1600 (2011).



\bibitem{Banfi:2012du} 
  A.~Banfi, M.~Dasgupta, S.~Marzani and L.~Tomlinson,
  Phys.\ Lett.\ B {\bf 715}, 152 (2012).



\bibitem{Aad:2012wfa} 
  G.~Aad {\it et al.} [ATLAS Collaboration],
  Phys.\ Lett.\ B {\bf 720}, 32 (2013).



\bibitem{Affolder:1999jh} 
  T.~Affolder {\it et al.} [CDF Collaboration],
  Phys.\ Rev.\ Lett.\  {\bf 84}, 845 (2000).



\bibitem{Abbott:1999wk} 
  B.~Abbott {\it et al.} [D0 Collaboration],
  Phys.\ Rev.\ D {\bf 61}, 032004 (2000).



\bibitem{Abazov:2010kn} 
  V.~M.~Abazov {\it et al.} [D0 Collaboration],
  Phys.\ Lett.\ B {\bf 693}, 522 (2010).



\bibitem{Khachatryan:2015pzs} 
  V.~Khachatryan {\it et al.} [CMS Collaboration],
  Phys.\ Lett.\ B {\bf 759}, 36 (2016).



\bibitem{Kang:2012am} 
  Z.~B.~Kang and J.~W.~Qiu,
  Phys.\ Lett.\ B {\bf 721}, 277 (2013).



\bibitem{Abazov:2007ac} 
  V.~M.~Abazov {\it et al.} [D0 Collaboration],
  Phys.\ Rev.\ Lett.\  {\bf 100}, 102002 (2008).



\bibitem{Aaij:2015gna} 
  R.~Aaij {\it et al.} [LHCb Collaboration],
  JHEP {\bf 1508}, 039 (2015).



\bibitem{Aaij:2016mgv} 
  R.~Aaij {\it et al.} [LHCb Collaboration],
  JHEP {\bf 1609}, 136 (2016).



\bibitem{Hirai:2007sx} 
  M.~Hirai, S.~Kumano and T.-H.~Nagai,
  Phys.\ Rev.\ C {\bf 76}, 065207 (2007).



\bibitem{deFlorian:2011fp} 
  D.~de Florian, R.~Sassot, P.~Zurita and M.~Stratmann,
  Phys.\ Rev.\ D {\bf 85}, 074028 (2012).



\bibitem{Kulagin:2014vsa} 
  S.~A.~Kulagin and R.~Petti,
  Phys.\ Rev.\ C {\bf 90}, no. 4, 045204 (2014).



\bibitem{Kovarik:2015cma} 
  K.~Kovarik {\it et al.},
  Phys.\ Rev.\ D {\bf 93}, no. 8, 085037 (2016).



\bibitem{Khanpour:2016pph} 
  H.~Khanpour and S.~Atashbar Tehrani,
  Phys.\ Rev.\ D {\bf 93}, no. 1, 014026 (2016).



\bibitem{Eskola:2016oht} 
  K.~J.~Eskola, P.~Paakkinen, H.~Paukkunen and C.~A.~Salgado,
  Eur.\ Phys.\ J.\ C {\bf 77}, no. 3, 163 (2017).



\end{thebibliography}
\end{document}